\documentstyle[galley,graphicx,epsf]{mn}
\begin{document}
\title {Photometric study of distant open clusters in the second quadrant: 
        NGC 7245, King 9, King 13 and IC 166}
                                                                                                                      
\author[Subramaniam \& Bhatt]
        {Annapurni Subramaniam$^{1}$, Bhuwan Chandra Bhatt$^{2}$
\thanks{email: purni@iiap.res.in (AS), bcb@crest.ernet.in}\\
$^{1}$  Indian Institute of Astrophysics, Bangalore 560034, India\\
$^{2}$  CREST Campus,  Indian Institute of Astrophysics, PO Box 19, Hosakote, Bangalore 562114, India}
                                                                                                                      
\maketitle
\label{firstpage}
\begin{abstract}
We present a UBV CCD photometric study of four open clusters, 
NGC 7245, King 9, IC 166 and King 13, located between
$l = 90^o - 135^o$. All are 
embedded in a rich galactic field. NGC 7245 and King 9 are 
close together in the sky and have similar reddenings.
The distances and ages are: NGC 7245, 3.8$\pm$0.35 kpc
and 400 Myr; King 9 (the most distant cluster in this quadrant) 7.9$\pm$1.1 kpc
and 3.0 Gyr. King 13 is 3.1$\pm$0.3 kpc distant and 300 Myr old.
King 9 and IC 166 (4.8$\pm$0.5 kpc distant \& 1 Gyr old) 
may be metal poor clusters (Z=0.008), as estimated from isochrone fitting.
The average value of the distance of young clusters from the galactic plane
in the above longitude range and beyond 2 kpc
($-$47$\pm$16 pc, for 64 clusters), indicates that the young disk bends 
towards the southern latitudes.

\end{abstract}
\section{Introduction}
Open star clusters are the best testing ground for stellar evolution,
galactic structure and chemical evolution of the disk. To increase the
sample of clusters beyond the 2 kpc radius in the second quadrant,
we observed four clusters, 
NGC 7245, King 9, King 13 and IC 166 with poorly known parameters. 

\section{Previous studies}
Yilmaz (1970) presented RGU photographic photometry of NGC 7245 
(Right Ascension (2000) 22:15:11, Declination (2000) +54:20:36; l=101.368, b= $-$1.852).
and obtained a
distance of 1925 pc and a reddening of 0.60 mag. WEBDA presents the UBV
photographic photometry by Karaali (1971) and CCD photometry by
Petry \& DeGioia-Eastwood (1994) for 67 stars. According to WEBDA,
the cluster is at a distance of 2106 pc, reddening is of 0.473 mag 
and log(age) is 8.246. Viskum et al. (1997)
performed BV CCD photometry 650 stars in a 7.7 $\times$ 7.7 arcmin$^2$ region
around NGC 7245 to detect $\delta$ Scuti stars. Using 
isochrone fitting, they estimated the age as 320 Myr, 
reddening of E(B-V) = 0.40$\pm$0.02 and distance as 2800 pc. 

None of the parameters is estimated for King 9 in the literature
(RA 22:15:30, Dec. +54:24:00). 
VI CCD photometry was performed by Phelps et al. (1994), but no parameters were 
estimated.
This is the first calibrated photometric study of King 9 in U and B bands and
the first estimation of cluster parameters.

Marx \& Lehmann (1979) obtained  UBV photographic photometry of
King 13 (RA 00:10:06, Dec. +61:10:00) and
estimated the distance as 1730 pc.
IC 166 (RA 01:52:30, Dec. +61:50:00) is a faint and distant cluster.
Burkhead (1969) obtained photographic
photometry of about 200 stars and photoelectric photometry of 20 stars 
in the central region.
Assuming a reddening of E(B-V)=0.8 mag, the distance was estimated as 3.3 kpc.
WEBDA presents the distance as 3970 pc, 
reddening as 1.05 mag and log(age) as 8.63.
From JHK-IR photometry, Vallenari et al. (2000) estimated the
age as $\sim$ 1 Gyr, reddening E(B-V) = 0.50 mag and
distance $\sim$ = 4.5 kpc.

\section{Observations}
Data were obtained using the 2-m Himalayan Chandra Telescope (HCT),
Hanle, IAO, operated by the Indian Institute of Astrophysics. The 
instrument used is the imaging camera of the Himalayan Faint Object Spectrograph and
Camera (HFOSC). The details of the telescope and the instrument are available 
from the web site, http://www.iiap.res.in/iao.html.
The CCD used for imaging is a 2 K $\times$ 4 K CCD,
where the central 2 K $\times$ 2 K pixels were used for imaging. The pixel size is 15 $\mu$
with an image scale of 0.297 arcsec/pixel. The total area observed is
approximately 10 $\times$ 10 arcmin$^2$.
The log of observations is given in Table 1. NGC 7245 and King 9 
are imaged in the same frame and hence they have the same log of observations.
The nights of observations were photometric and Landolt standards were
observed for photometric calibrations. The seeing was between 1 -- 1.5 arcsec.
The initial CCD reductions were carried out using IRAF. 
DAOPHOT II (Stetson 1987) was used for magnitude estimation and calibration.
The transformation equations used are:\\
$u = U + 3.447 -0.097 (U-B) + 0.327X $\\
$b = B + 1.190 + 0.049 (B-V) + 0.180X $\\
$v=V + 0.725 - 0.060 (B-V) + 0.094X$\\
The zero point errors are 0.010, 0.013 and 0.018 magnitudes in V, B and U respectively.

We compared our photometry of NGC 7245 with that of Petry \& DeGioia-Eastwood (1994).
We could identify 48 common stars. 
The average and
the standard deviation of the difference (our data $-$ other data) 
is 0.003$\pm$0.025 in V magnitude and $-$0.04$\pm$0.09 in (B$-$V) magnitude.
The V band photometric data of King 9 were compared with those of Phelps et al. (1994).
297 stars were found to be common. The data were found to compare well for most of the stars, except for
some outliers. If we ignore the outliers, the average value of the
difference is 0.06$\pm$0.08 in V magnitude. 

\renewcommand{\thetable}{1}
\begin{table*}
\centering
\caption{Log of photometric observations}
\begin{tabular}{lrrr}
\hline
Cluster& Date & Filter & Exp time (sec) \\
\hline
NGC 7245/ King 9 & 19 September 2003&  V & 1, 10, 60, 120\\
      &  &  B & 10, 30, 300\\
      & &  U & 60, 2X300\\
IC 166 & 19 September 2003&  V & 10, 60, 120\\
      &  &  B & 30, 180, 300\\
      & &  U & 180, 600\\
King 13 & 18 September 2003&  V & 5, 30, 60\\
      &  &  B & 10, 60, 300\\
      & &  U & 180, 600\\
\hline
\end{tabular}
\end{table*}

\section{NGC 7245}
%
%
\begin{figure}
\epsfxsize=7truecm
\epsffile{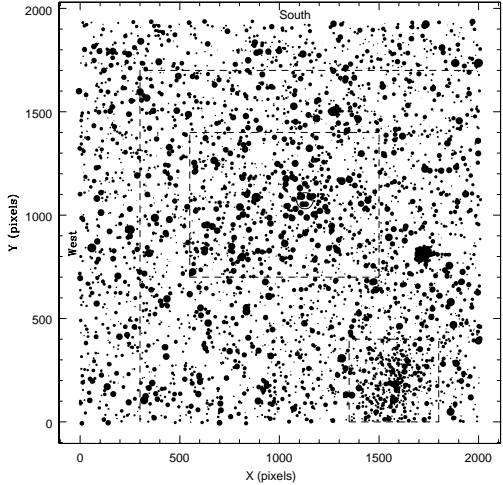}
\caption{The observed region of the cluster NGC 7245. 
The cluster area of NGC 7245,  King 9 (lower right side) and
field regions considered as control fields are indicated by dashed lines.}
\end{figure}

%
%
\begin{figure}
\epsfxsize=7truecm
\epsffile{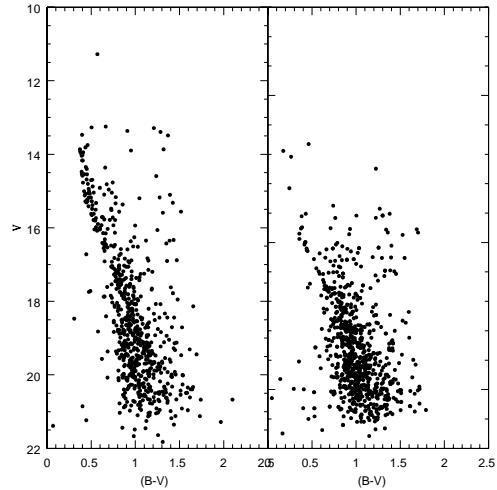}
\caption{The CMDs of the cluster (left) and the field region (right) of NGC 7245.
}
\end{figure}

%
%
\begin{figure}
\epsfxsize=7truecm
\epsffile{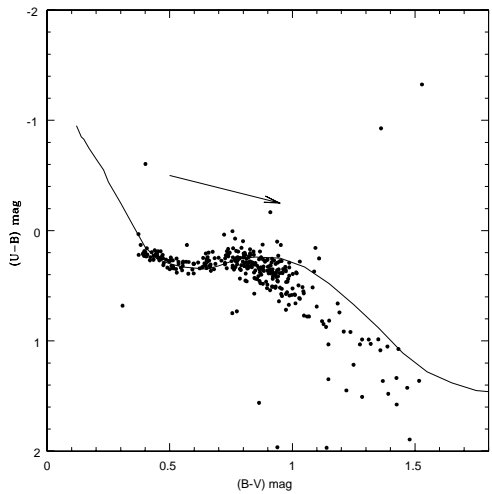}
\caption{Estimation of reddening towards NGC 7245 using the (U$-$B) vs (B$-$V) diagram. 
} 
\end{figure}
NGC 7245 is a sparse cluster (figure 1), without a well defined center.
The adopted cluster area is indicated by dashed lines and the center by an open circle. 
A number of bright
stars was found to the left of the cluster. These stars were found to delineate the
upper main sequence (MS). Thus the cluster region is elongated to the left to
include these bright stars.
King 9 is found as a faint and condensed cluster at the lower right side
of the field. 
The adopted cluster area for this cluster is also shown by dashed lines.
In order to delineate the cluster sequence from  contamination by field stars,
we defined a field region close to the periphery of the observed region.

The colour-magnitude diagram (CMD) of the cluster region
shows a well defined MS as compared to that of the field region (figure 2).
(U$-$B) vs (B$-$V) colour-colour diagram of stars inside the cluster
region (figure 3), is used to estimate the reddening.
The fit is good for the blue side, whereas it fails in the red side. Since the MS of
the cluster lies in the blue side, we fit the blue side rather than the red side.
Estimated value of the reddening is 0.45$\pm$0.02 mag. 

%
\begin{figure}
\epsfxsize=7truecm
\epsffile{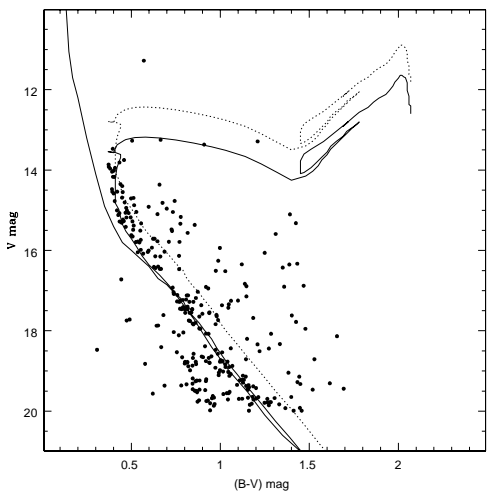}
\caption{400 Myr isochrone (continuous line) is fitted to the field subtracted 
CMD of NGC 7245. The ZAMS is also over plotted on the CMDs. 
Dotted lines correspond to the isochrone for binary stars with mass ratio equal to 1.
}
\end{figure}

To identify the cluster sequence clearly, the field stars must
be removed from the cluster CMD. The brighter stars 
which do not share the
cluster reddening are removed. For fainter stars, 
we statistically subtracted the 
field stars from the cluster CMD. The data can be assumed to be complete up to V = 19 mag.
Since the cluster and the field regions have the same area, the stars in the field CMD
can be used to statistically remove the field stars.
This is done using the zapping technique, where the candidate field stars in the cluster
CMD are removed by cross-correlating the cluster CMD with the field CMD.
Maximum search size used is 0.2 mag in V and 0.1 mag in (B$-$V). In the case of NGC 7245,
half of the above values were used.
After this procedure, the cluster sequence gets delineated very clearly (figure 4). 
Most of the field stars found to the right side of the MS are removed. 
On the other hand, there are some stars located to the left of the MS and 
fainter than V=18 mag.
We consider the MS which is brighter than
V=18 mag for the estimation of cluster parameters. ZAMS is fitted to the cluster MS 
to estimate a distance modulus as (V$-M_V$)$_0$ = 12.9$\pm$0.2 mag (distance = 
3800$\pm$350 pc). The above estimate
is also supported by the isochrone fits to the cluster sequence.
We used Bertelli et al. (1994) isochrones for solar metallicity. The age of the cluster is
log (age) = 8.6, which corresponds to 400 Myr. The isochrone shown as a bold
line is for 400 Myr and the isochrone for the binary stars is shown as a dotted line.
The evolved part of the CMD is well fitted by the isochrone. 
In the MS, the
stars are uniformly populated between the single and the binary isochrone indicating a
healthy population of binaries. There is a strong indication of a MS gap at 
(B$-$V) $\sim$ 0.7 mag, which corresponds to (B$-$V)$_0$ $\sim$ 0.25 mag.
This could be the Bohm-Vitense gap at (B$-$V)$_0$ = 0.25 mag. 
This
gap is likely to be a real feature, as it is seen even before the field
star subtraction (see figure 2). 

\section{King 9}

%
%
\begin{figure}
\epsfxsize=7truecm
\epsffile{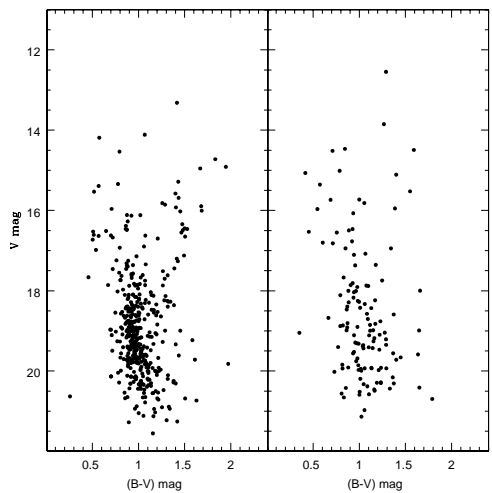}
\caption{The CMD of the cluster region (left) and field region (right) of King 9.
}
\end{figure}

%
%
\begin{figure}
\epsfxsize=7truecm
\epsffile{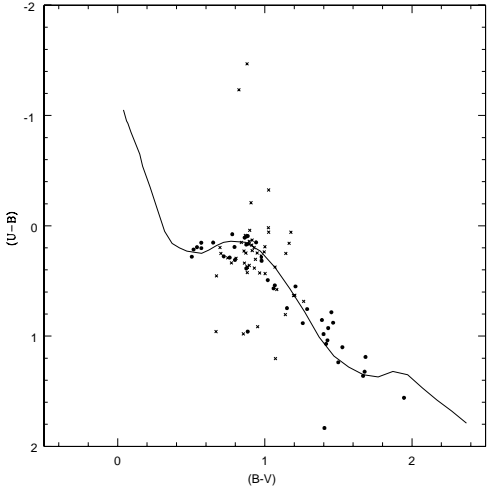}
\caption{Estimation of reddening towards King 9 using the (U$-$B) vs (B$-$V) diagram. 
Crosses indicate stars brighter than V=17.5 (likely to be field stars) and dots indicate
stars fainter than V=17.5 mag (potential cluster members).}
\end{figure}

%
%
\begin{figure}
\epsfxsize=7truecm
\epsffile{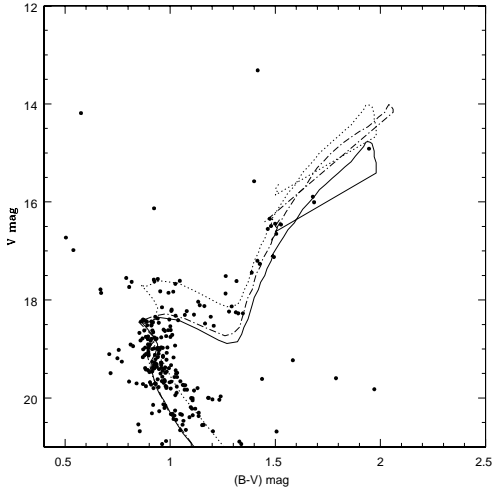}
\caption{The field star subtracted CMD is fitted with solar metallicity isochrone of
log (age) 9.5, shown as a bold line. The dotted line indicates the binary isochrone
for a mass ratio of 1.0.
Dashed line indicates Z=0.008 isochrone of the same age. Notice that the red clump is
better fitted by this isochrone. }
\end{figure}

King 9 is a compact cluster found very close to NGC 7245 in the sky. It appears to be
farther and older than NGC 7245.
A field area of equal size is selected from the left bottom side.
In the cluster CMD (left panel, figure 5), the MS is clearly visible, in sharp
contrast to the field CMD (right panel, figure 5). Due to the faintness of the cluster,
only 70 stars were detected in the U band and these stars are used to
estimate the reddening (figure 6). The
reddening is E(B$-$V) = 0.37 $\pm$ 0.04, 
similar to the value found for NGC 7245. The large error in the reddening
is due to the small number of cluster stars. The turn-off of the cluster MS is
at $\sim$ 18.0 mag and hence ZAMS fitting could not be used to estimate the
distance. We eliminated the field stars in the
King 9 cluster CMD, using a procedure similar to that adopted for NGC 7245. 
In the cleaned CMD (figure 7), MS and turn-off are very clearly visible,
even though it is just 2 magnitudes brighter than the limit of detection.
The isochrone fitting to the cluster CMD
is also shown. The age of the cluster is log(age) = 9.5  (3.0 Gyr)
and the distance modulus is 14.5$\pm$ 0.3 mag (distance =7.9$\pm$1.1 kpc). 
Accordingly, we find that King 9 is the
farthest open cluster in the second quadrant. The 3.0 Gyr isochrone for
solar metallicity does not match the giant branch very well. It is too red to fit
the observed stars. We have also shown the 
binary sequence (dotted isochrone), where the giant branch is fitted well. 
This sequence fits the
stars brighter than the MS very well, indicating that the cluster may have a significant
fraction of binary population including some blue stragglers. 
The isochrones for Z=0.008 (dashed lines)
fits the red clump and the giant branch well, for the same age. 
Therefore, this distant cluster could be
metal poor. The metal poor isochrone estimates a reddening of 0.47$\pm$0.04 mag 
and a distance of 7.0$\pm$1.1 kpc.

\section{King 13}
%
%
\begin{figure}
\epsfxsize=7truecm
\epsffile{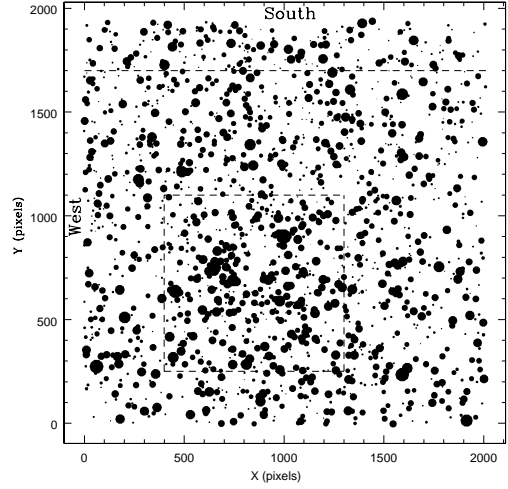}
\caption{The observed field of King 13 along with the location of cluster
and field areas.}
\end{figure}
%
%
\begin{figure}
\epsfxsize=7truecm
\epsffile{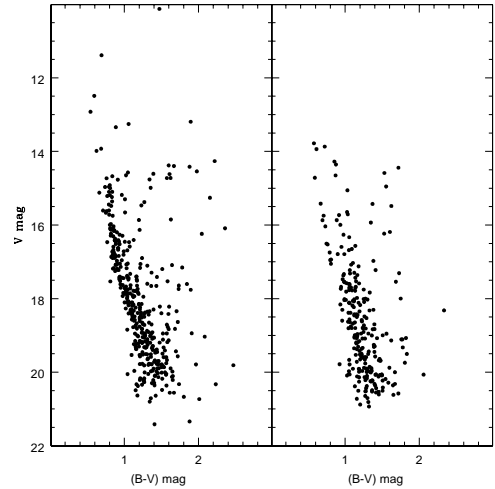}
\caption{Observed CMD of the cluster region  (left) and field region (right) of King 13
}
\end{figure}
%
%
\begin{figure}
\epsfxsize=7truecm
\epsffile{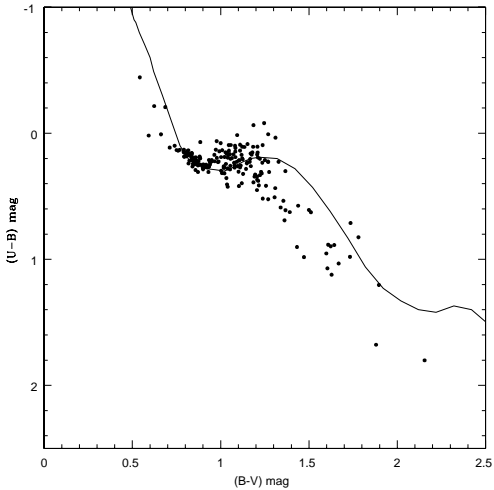}
\caption{Estimation of reddening towards King 13 using the (U$-$B) vs (B$-$V) diagram.} 
\end{figure}
%
%
\begin{figure}
\epsfxsize=7truecm
\epsffile{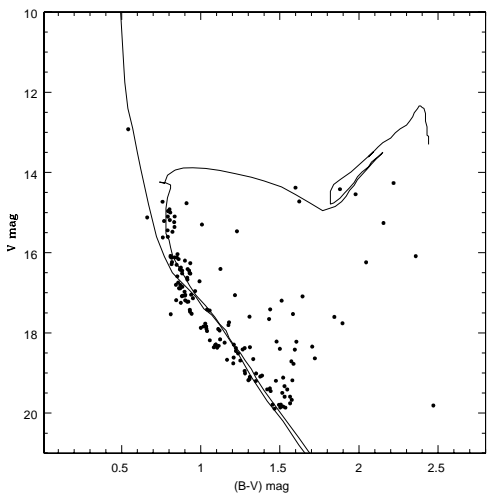}
\caption{An age of (log(age) = 8.5) is estimated for King 13 using isochrone fit.
}
\end{figure}

King 13 (figure 8) does not have
a well defined center, but a density enhanced ring can be seen. An estimate of the
cluster region is shown as a dotted line, which is about 2 arcmin from the visually estimated
cluster center. 
The CMDs of the cluster and the field regions (figure 9, left and
right panels respectively) show a clear MS in the cluster CMD.
The fit of the reddening curve (figure 10) is merely satisfactory,
 as not many young and blue stars
are found to fit the curve properly. The reddening 
is 0.82 $\pm$ 0.02 mag. The cleaned CMD of the cluster region is shown in
figure 11, where the ZAMS fit and the isochrone fits are also shown. The
distance modulus is 12.5 $\pm$0.2 mag (distance = 3100 $\pm$330 pc).
The age of the cluster is log(age) = 8.5 (age = 300 Myr). 

\section{IC 166}
%
%
\begin{figure}
\epsfxsize=7truecm
\epsffile{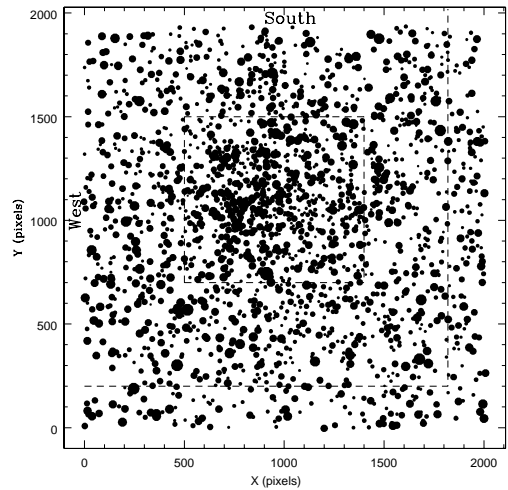}
\caption{Observed region of IC 166 where cluster and field locations are demarcated.}
\end{figure}
%
%
\begin{figure}
\epsfxsize=7truecm
\epsffile{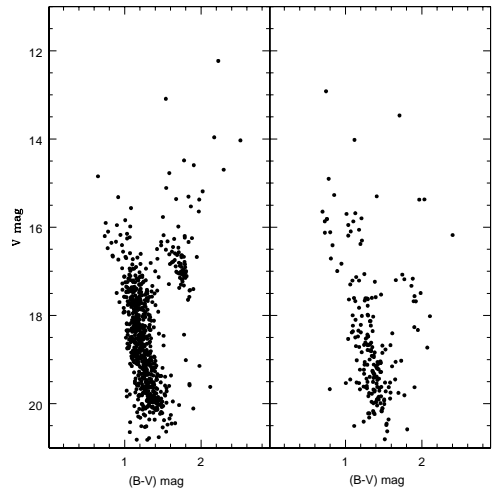}
\caption{The cluster (left) and field (right) CMD of IC 166
}
\end{figure}
%
%
\begin{figure}
\epsfxsize=7truecm
\epsffile{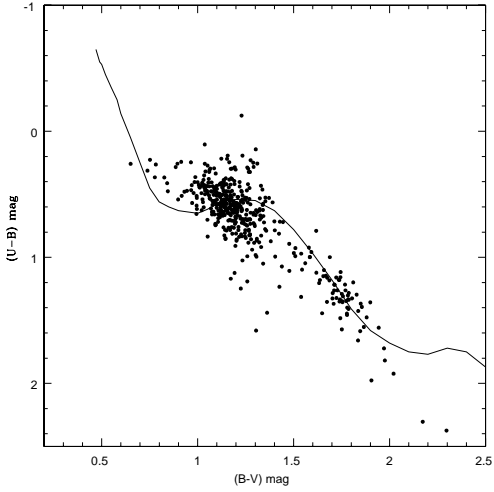}
\caption{Estimation of reddening towards IC 166 using the (U$-$B) vs (B$-$V) diagram.} 
\end{figure}
%
%
\begin{figure}
\epsfxsize=7truecm
\epsffile{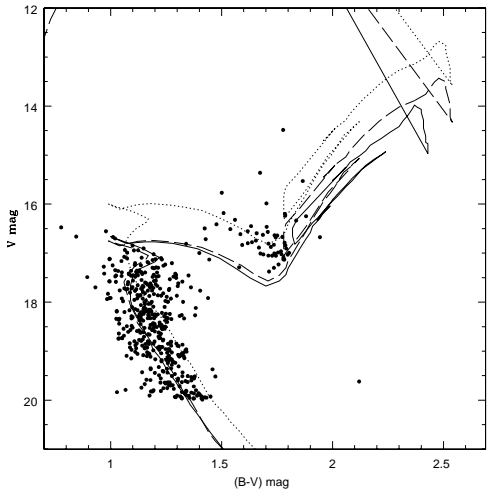}
\caption{The field star subtracted CMD is fitted with solar metallicity isochrone of
log (age) 9.0, shown as a bold line. The dotted line indicates the binary isochrone
for a mass ratio of 1.0.
Dashed line indicates Z=0.008 isochrone of the same age. Notice that the red clump is
better fitted by this isochrone. }
\end{figure}

IC 166 is a rich cluster (figure 12) and the  
center of the cluster is at X=917 and Y=1058. From the radial density profile,
the cluster radius is estimated as 4 arcmin. We consider the central
condensed region of the cluster within the radius of 2.5 arcmin to estimate the parameters.
The region considered as field is also shown.  The
cluster sequence is clearly visible (left panel, figure 13), along with a significantly
elongated red clump. The CMD of the field region (right panel, figure 13) 
shows a scattered sequence, with a number
of bright stars, which are also seen in the cluster CMD. 
The reddening sequence is fitted to the red clump stars (figure 14), and the reddening is
E(B$-$V) = 0.80 $\pm$ 0.02 mag. The fit for the red clump stars is good, whereas it is not
so good for the MS stars. The field star corrected CMD (figure 15) shows that the
brighter stars in the CMD are removed, and the turn-off is clearly visible. Isochrone
for an age of 1.0 Gyr is shown as a continuous line and its binary isochrone as a dotted line.
This isochrone fits the turn-off very well. On the other hand,
it fails to fit the red clump stars. Some of the brighter subgiants could be binaries as they
fall on the binary isochrone. In order to fit the red clump stars, isochrone of lower metallicity
(Z=0.008) is also plotted (dashed line). The fit of this isochrone to the red clump stars
is better, as it extends to fainter magnitudes. 
For a better fit, the isochrone needs to reach fainter magnitudes. This
gives an indication that the cluster could be more metal poor than Z=0.008. 
The fit shown for solar metallicity isochrone
indicates the following cluster parameters: reddening E(B$-$V)= 0.80 $\pm$0.02, 
distance modulus = 13.4$\pm$0.2, (distance = 4.8$\pm$0.5 kpc) and an age of 1 Gyr.  
The fit shown for
Z=0.008 is obtained for the following cluster parameters: reddening E(B$-$V)= 0.92 $\pm$0.02, 
distance modulus = 13.2$\pm$0.2, (distance = 4.2$\pm$0.5 kpc) and an age of 1 Gyr.   
Friel \& Janes (1993) estimated the metallicity of IC 166 as [Fe/H] = $-$0.32$\pm$0.20, 
which was later
revised to $-$0.41. This supports the conclusion that the cluster is metal poor.

\section{Structure of the outer galactic disk between $l= 90^o- 135^o$ }
\begin{figure}
\epsfxsize=7truecm
\epsffile{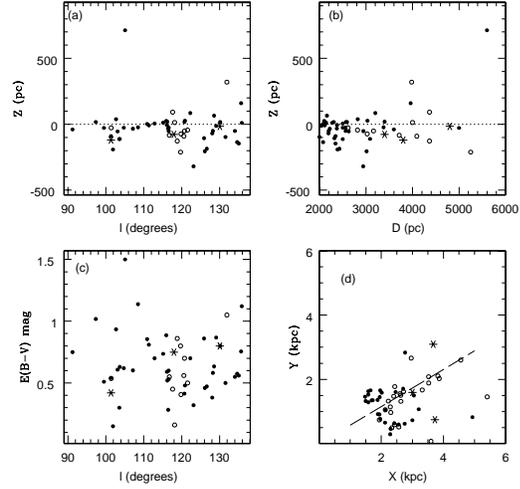}
\caption{Distribution of clusters between $l=90^o - 135^o$ and beyond 2 kpc.
Figure (a):
Distribution of clusters in Z (pc) as a function of the galactic longitude, $l$. Clusters younger than 100 Myr are shown as dots and clusters older 
than 100 Myr and younger than 1 Gyr are shown as open circles. 
Figure (b): Their distribution in Z (pc) as a function of Distance D (pc).
Figure (c): Reddening distribution as a function of longitude, $l$.
Figure (d): Cluster distribution in the galactic coordinates.
This figure is meant to trace any spiral beyond Perseus spiral arm.  The dotted line shows the 
location of $l=130^o$.}
\end{figure}
We have estimated the basic parameters of 4 poorly studied open clusters found in the second
galactic quadrant, between $l$ = 90$^o$ - 135$^o$. Three clusters are at a distance of about 
4 kpc and one cluster is beyond 7 kpc.
Here, we try to find how these clusters have contributed to the understanding of
the galactic disk in this direction, beyond 2 kpc. We have compiled the parameters of clusters
in this part of the outer disk.
The clusters 
studied here are shown as asterisks (figure 16).  Clusters older than 1 Gyr are not shown in the figure. 
Therefore, the location of King 9 is also not shown, as it is
older than 1 Gyr and at a large distance. This cluster is located about 240 pc below the plane. 
Figure 16(a)
shows the distribution of clusters between $l=90^o - 135^o$, as a function of their distance
from the plane (Z). Figure 16(b) shows the distribution of their distance as a function of Z.
The dots indicate clusters younger than 100 Myr and open circles indicate clusters older than
100 Myr, but younger than 1.0 Gyr.
It is clear from these plots that the majority of clusters are found below the
galactic plane. One cluster, Be 93 is found at a distance of 700 pc to the north.
Be 93 is just 100 Myr old. Be 93 was not considered for the calculation mentioned below.
It will be interesting to study the kinematics of Be 93, in future.
We estimated the average value 
of Z from our sample. 19 (32\%) clusters are found above and 41 (68\%) clusters 
are found below the galactic equator.
The average value of Z is $-$47$\pm$16 pc. Thus
the galactic disk, as delineated by clusters, is bent to the south.
All the four clusters studied
in this paper are in the southern part of the disk. From figure 16(b), it is clear
that these clusters have increased the number of clusters known beyond 3.0 kpc.
This figure also indicates that the structure of the disk beyond 4.0 kpc is not
clearly seen. 

This small deviation of the galactic disk presents to the 
southern latitudes may be considered as a mild warp in the young disk.
Momany et al. (2006) derived
stellar warp using red clump stars located between 2 - 4 kpc from the Sun 
(their figure 8). It can be seen that, the disk shows a marginal 
bend towards the negative latitudes.
The cluster distribution might be indicating the same bend in the 
disk at these longitudes and
at similar distances from the Sun. Thus the young disk follows the warp 
of the intermediate age red clump stars.
It will
be interesting to see the young disk beyond 4.0 kpc, as this kink seen in red 
clump stars disappears
at larger distances. In fact, the disk bends towards the northern 
latitudes (Momany et al. 2006,
figure 8). Thus it is essential to increase the sample of young clusters beyond 
4.0 kpc in this quadrant. Pandey et al. (2006) found a similar warp between $l=100-130$
when they traced the background population of open clusters
in this direction (their figure 6).

The average value of
reddening increases from 90$^o$ towards 135$^o$ (figure 16(c)).
The reddening estimates of the clusters
studied here are in agreement with those estimated for other nearby clusters. 
The Perseus arm stretches
between 2 - 3 kpc (figure 16(d)), which is the locus of dots 
representing open clusters younger than
30 Myr (equivalent to log (age)=7.5).
At $l \sim 130^o$, there is a stretch of clusters outward
up to a distance of $\sim$ 5.5 kpc. The clusters found in this extension are all older than
30 Myr. Hence this cannot be considered as an extension of the Perseus spiral arm.
This figure does not support the existence of a spiral arm
beyond the Perseus arm. Two of the clusters, NGC 7245 and IC 166, are found 
beyond the Perseus arm, while the cluster King 13 is part of the outward stretch.
Pandey et al. (2006) have shown a similar extension in their figure 6, as delineated by
the background population towards open clusters.

King 9 and IC 166 are located beyond a 
galactocentric radius of 10 Kpc. According to the
observed radial abundance gradient of open clusters (Friel et al. 2002, figure 2), these
clusters are expected to be metal poor ($[Fe/H] \sim -0.3$). 
King 9 \& IC 166 also belong to an older group, as old as or older than 1 Gyr.
These two
clusters are important targets to derive the metallicity gradient in the disk as a
function of radial distance.
There are only 4 known open clusters, which are older than log(age) = 9.4 and located
beyond 7.0 kpc. Three of them are found in the third quadrant and one in the fourth quadrant.
King 9, the farthest old open cluster in the second quadrant,
is a potential target to study the kinematics and abundances to understand the properties
of the galactic disk at its extremes.

\section {Acknowledgments}
We thank the staff of Indian Astronomical Observatory, Hanle-Ladakh and at CREST Campus, Hosakote for assistance during observations. This research has made use of WEBDA database, operated at the Institute of Astrophysics, University of Vienna.

{}

\begin{thebibliography}{}
\bibitem[1994]{b94}Bertelli, G., Bressan, A., Chiosi, C., Fagotto, F., Nasi, E., 1994, A\&AS, 106, 275
\bibitem{} Burkhead, M.S., 1969, AJ, 74, 1171
\bibitem{} Friel, E.D., Janes, K.A., 1993, A\&A, 267, 75
\bibitem{} Friel, E.D., Janes, K.A., Tavarez, M., Scott, J., Katsanis, R., Lotz, J., Hong, L., Miller, N., 2002, AJ, 124, 2693
\bibitem{} Karaali S., 1971, Publ. Istanbul. Univ. No. 92
\bibitem{} Marx, S., Lehmann, H., 1979, Astron. Nach., 300, 295 
\bibitem{} Momany, Y., Zaggia, S., Gilmore, G., Piotto, G., Carraro, G., Bedin, L. R., de Angeli, F., 2006, A\&A, 451, 515
\bibitem{} Pandey, A.K., Sharma, S., Ogura, K., 2006, MNRAS, 373, 255
\bibitem{} Phelps, R.L., Janes, K.A., Montgomery, K.A., 1994, AJ, 107,1079
\bibitem{} Petry C.E., DeGioia-Eastwood K., 1994, priv. comm.
\bibitem{} Stetson, P.B., 1987, PASP, 99, 191
\bibitem{} Vallenari, A., Carraro, G., Richichi, A., 2000, A\&A, 353, 147
\bibitem{} Viskum, M., Hernandez, M. M., Belmonte, J. A., Frandsen, S., 1997, A\&A, 328, 158
\bibitem{} Yilmaz, F., 1970, A\&A, 8, 213

\end{thebibliography}
\end{document}